\documentclass[showpacs,twocolumn,amssymb,aps,flushbottom,floatfix,balancelastpage,superscriptaddress]{revtex4-2}
\usepackage{graphicx}
\usepackage{amsmath}
\usepackage[colorlinks=true]{hyperref}
\usepackage{subfig}
\usepackage{natbib}
\usepackage{enumitem}

\usepackage{mathtools}
\begin{document}

\title{From rogue waves to solitons}
\author{Amdad Chowdury}
\affiliation{School of Physical and Mathematical Sciences, Nanyang Technological University, 637371, Singapore}

\author{Wonkeun Chang}
\affiliation{School of Electrical and Electronic Engineering, Nanyang Technological University, 639798, Singapore\looseness=-1}

\author{Marco Battiato}
\affiliation{School of Physical and Mathematical Sciences, Nanyang Technological University, 637371, Singapore}

\begin{abstract}
Using a generalized nonlinear Schr\"odinger equation, we investigate the transformation of a fundamental rogue wave to a collection of solitons. Taking the third-order dispersion, self-steepening, and Raman-induced self-frequency shift as the generalizing effects, we systematically observe how a fundamental rogue wave has an impact on its surrounding continuous wave background and reshapes its own characteristics while a group of solitons are created. We show that under the influence of the self-steepening effect, a finite-volume rogue wave can transform into an infinite volume soliton. Also, we find that with the Raman-induced self-frequency shift, a decelerating rogue wave generates a red-shifted Raman radiation while the rogue wave itself turns into a slow- moving soliton. We show that each of these effects has an element of mechanism that favors the rogue wave to generate a group of solitons while the rogue wave itself also becomes one of these solitons.
\end{abstract}
\maketitle

\section{Introduction}
\label{L1}
After the discovery of a rogue wave solution in nonlinear Schr\"odinger equation (NLSE) by D.~H.~Peregrine in 1984, there was a surge of interest because of its ability to explain formation of a sudden giant oceanic wave, which is often called a rogue wave or a monster of the deep \cite{Dysthe:08:ARFM}. While at the beginning, most of the investigations were focused on applying the solution to oceanic rogue waves \cite{Akhmediev:09:PLA1,onorato2001freak,muller2005rogue}, the idea has gradually expands to other fields such as nonlinear optics \cite{Kibler:10:Nature}, plasma physics \cite{moslem2011surface}, atmospheric science \cite{stenflo2010rogue}, superfluids \cite{Efimov:10:EPJST}, Bose-Einstein condensates \cite{Bludov:10:EPJST}, capillary waves \cite{Shats:10:PRL}, acoustic waves \cite{moslem2011dust,tsai2016generation}, electromagnetic waves \cite{veldes2013electromagnetic}, matter waves \cite{bludov2009matter}, and economics \cite{zhen2010financial}. 

Among many systems where the concept of rogue wave formation is materialized, hydrodynamics and optics are the two areas where most of new discoveries are made. This is because both water and optical fiber posses dispersive and nonlinear properties that can be modeled by the NLSE. The presence of rogue waves in both of these media is confirmed experimentally \cite{Solli:07:Nature,Kibler:10:Nature,Chabchoub:11:PRL}. Thus, rogue wave observations can now be made in a water-wave tank or on an optical table. In particular, due to wide availability of various optical components, the research in optical rogue wave became popular since its first discovery 14 years ago \cite{Solli:07:Nature}. A comprehensive overview of the recent progress in optical rogue wave research can be found in \cite{akhmediev2013recent,song2020recent}. 

The basic form of NLSE has limitations when modeling ultrashort pulse propagation as it can trigger higher-order linear and nonlinear optical effects \cite{agrawal2011nonlinear,Agrawal:12:Book}. Among them, the third-order dispersion (TOD), self-steepening (SS), and Raman-induced self-frequency shift (RIFS) are the most dominant mechanisms that can directly affect the pulse. The inclusion of these terms in the NLSE destroys its integrability, and hence a majority of theoretical studies of their effect on optical rogue waves so far have been carried out either numerically or by taking only one or two effects at a time. The result is an incomplete picture of the rogue wave dynamics under these higher-order effects. Moreover, many studies focus only on how these effects impact the central structure of the rogue wave solutions or whether the solutions can survive the perturbations \cite{ankiewicz2009rogue,bandelow2012persistence,ankiewicz2013rogue-even,ankiewicz2013rogue}, and little efforts have been made to understand their impact on the neighboring continuous wave background. The modulation instability (MI) which generates rogue wave-like structures creates a variety of other sub-structures originating from this continuous wave background. This is apprarent in a MI-based supercontinuum generation, where a long pulse undergoes MI, and the end product is a shower of hundreds of fundamental solitons \cite{russell2014hollow,travers2011ultrafast}. How these soliton bunches are formed in the midst of MI has not yet been clearly described.

In this work, we conduct a comprehensive study on the temporal and spectral properties of an optical rogue wave in the generalized nonlinear Schr\"odinger equation (GNLSE) that includes the TOD, SS, and RIFS effects. By first taking individual effects separately, we study how each one impacts the rogue wave solution in the temporal and spectral domains {\it at the time of the rogue wave appearance}. We also present what changes these three effects produce on the surrounding waves {\it after the emergence of the rogue wave} and the fate of the rogue wave after it evolves for a long time. Finally, we apply the three effects simultaneously and observe the combined evolution characteristics. We find that a fundamental rogue wave triggers a collection of solitons from its emerging point while the rogue wave itself also transforms into a soliton under the influence of TOD, SS, and RIFS effects.

\subsection{Model, solution, and techniques}

The GNLSE in its normalized form is \cite{agrawal2011nonlinear}:
\begin{multline}
\label{eq2}
i \frac{\partial \psi}{\partial z}-\frac{\beta_2}{2} \frac{\partial^2 \psi}{\partial t^2}+\gamma \, \psi\lvert\psi\rvert^2=\\
i\epsilon_3\, \frac{\partial^3 \psi}{\partial t^3}-i s\,\frac{\partial}{\partial t}(\psi\lvert\psi\rvert^2)+\tau_R \psi \frac{\partial|\psi|^2}{\partial t}\textrm{,}
\end{multline}
where $\beta_2$, $\gamma$, $\epsilon_3$, $s$, and $\tau_R$ are the normalized coefficients of the group-velocity dispersion, optical Kerr effect, TOD, SS, and RIFS, respectively. The explicit expressions of the coefficients in Eq.~(1) are:
\[
\begin{array}{ccc}
\epsilon_3& =  & \frac{\beta_3}{6|\beta_2|\,t_0}, s=\frac{1}{\omega_0\,t_0},\tau_R=\frac{T_r}{t_0},
\end{array}
\]
where $\omega_0$ is the carrier angular frequency, $t_0$ is the pulse duration, $\beta_3$ the TOD parameter, and $T_r$ is the Raman time constant \cite{atieh1999measuring}. The TOD, SS, and RIFS effects are inversely proportional to the pulse duration $t_0$, i.e.~, their contributions can be negligible when $t_0$ is large, or significant when it is small. Setting $\epsilon_3=s=\tau_R=0$, $\beta_2=-1$, and $\gamma=1$, Eq.~(\ref{eq2}) becomes NLSE, which can be solved analytically using the inverse scattering transformation (IST) \cite{Zakharov:72:JETP}.

We assume that TOD, SS, and RIFS are the perturbations to the NLSE. We take small values of these effects and study their impact on the rogue wave dynamics. To find numerical solutions in the perturbed system, we pick an initial condition from the exact analytical first-order rogue wave solution of NLSE, which is given as:  
\begin{equation}
\label{eq3}
\begin{aligned}
\psi(z,t)&= \left[1-\frac{G+ i H (z-z_{1})}{D}\right] e^{i~(z-z_{1})}\textrm{.}
\end{aligned}
\end{equation}
The solution can be derived by applying the Darboux transformation technique using the plane wave $\psi=\text{exp(iz)}$ as an initial seed solution where $G=4$, $H=8$, and $D=1+4t^2+4(z-z_{1})^2$ \cite{Akhmediev:09:PRE}. The initial condition in our study should start from practically a minuscule amplitude that works as an infinitesimally small modulation on the background. Therefore, we use the solution well before it emerges into a full-height rogue wave; i.e., we set $z_{1} = 30$. These effects exponentially amplify the small modulation which eventually appears as a rogue wave solution. The details of the numerical techniques applied to solve the perturbed NLSE-type equation is provided in \cite{chowdury2021rogue}.

In a chaotic wave field, various types of wave coexist together such as plane waves, solitons, and breathers. The IST technique is a useful mean to classify a random wave field into these types. We use it to indentify the types of waves formed on top of the background wave when NLSE is perturbed. This involves finding the eigenvalues of a given potential as a particular wave form, which provides a spectral portraits of soliton-type solutions. The IST spectral problem is generally solved in the framework of finite-gap theory \cite{randoux2016inverse}. Depending on a given IST spectral portrait, the genus, $g$, identifies what type of solution the spectrum belongs to. $g$ is measured as $J-1$ where $J$ is the number of spectral bands in the given IST spectrum. The rogue wave type solutions have $g=2$ as they have three spectral bands. The plane waves have $g=0$, while the soliton solutions have $g=1$. The details of the use of IST techniques to classify various localized wave forms are discussed in \cite{randoux2016inverse,randoux2018nonlinear,bonnefoy2020modulational}.

\section{Effect of third-order dispersion}
\label{L3}
Taking the effect of TOD only, the generalized NLSE takes the form:
\begin{equation}
\label{eq4}
i \frac{\partial \psi}{\partial z}+\frac{1}{2} \frac{\partial^2 \psi}{\partial t^3}+\psi\lvert\psi\rvert^2-i\epsilon_3\, \frac{\partial^3 \psi}{\partial t^3}=0\textrm{.}
\end{equation}
\begin{figure}[htbp]
\centering
\includegraphics{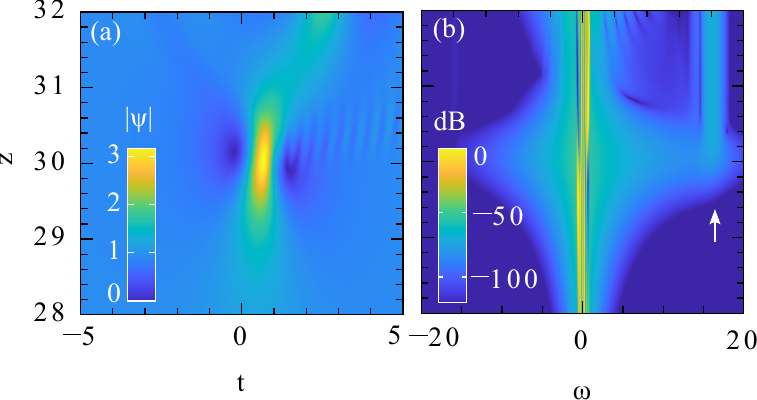} 
\caption{(a) Temporal amplitude evolution of the emergence of a first-order rogue wave at $z=30$ with $\epsilon_3=-0.03$ (b) Corresponding spectral intensity evolution. It shows the emission of dispersive wave shown by the white arrow as observed in Ref.~\cite{baronio2020resonant}. Note that in analytic solution the rogue wave appears at $z=0$ where as in simulation it is at $z=30$.}
\label{fig2}
\end{figure}

No analytic rogue wave solutions exist for the NLSE with the TOD term. Numerical descriptions of the rogue wave solutions in wave turbulence have been presented in \cite{kibler2011rogue,taki2010third}.  Similarly, the dispersive wave emission from a rogue wave as a result of TOD, shown in Fig.~\ref{fig2} indicated by the white arrow, has been demonstrated in \cite{baronio2020resonant}. The rogue wave emerges at around $z=30$ in Fig.~\ref{fig2} with a drift velocity while radiating a phase-matched dispersive wave following the relation $\omega_{DW}=3/\beta_3$. Similar type of radiations also occur when higher-order soliton is perturbed by TOD \cite{Cherenkov}. 

\begin{figure}[ht]
\begin{center}
\includegraphics{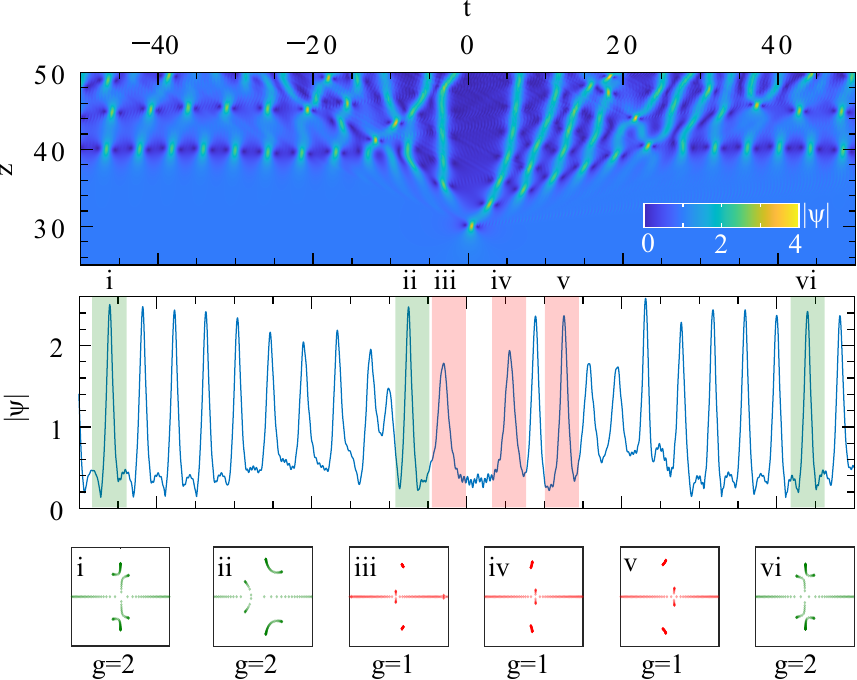}
\caption{(Top panel) Temporal amplitude evolution of a rogue wave under the influence of TOD. The rogue wave appears at $z=30$, and a skewed background wave dynamics is visible with a combination of different types of wave entities. (Mid-panel) The wave envelop is an instance extracted at $z=40$ after the emergence of the rogue wave. (Bottom panel) IST spectra of various types of wave profiles. }
 \label{fig3}
\end{center}
\end{figure}

TOD has a significant influence on the surrounding wave environment as well as reshaping the rogue wave while it evolve for a long time. Our earlier study \cite{chowdury2021rogue} has indicated that TOD transforms a rogue wave eventually to a collection of solitons. It is the lateral evolution of the rogue wave after its development that gives rise to the soliton-type waves. For instance, as shown in Fig.~\ref{fig3} (top panel), the background waves are creating a host of various other types of waves beyond the point of the rogue wave emergence at $z=30$ as a result of TOD.

To identify the types of wave generated, we use the IST technique for spectral analyses. We take an instances of the solution at $z=40$ after the emergence of the rogue wave in Fig.~\ref{fig3}, and choose six localized formations from the profile presented in the mid-panel. There are two types of wave formed, breather and soliton-type which are grouped by green and red shaded area. As indicated in the insets (bottom-panel), the entire transverse profile reveals IST spectra of solitons with the appearance of breathers and new rogue waves. Note that the rogue wave is also a class of breather-type solutions. The breather or $g=2$ solution arises near the edges of the field showing three distinct IST-spectral bands in insets i, ii, and vi. Solitons arise in the middle of the profile, where they are ejected from near the emerging point of the rogue wave at $z=30$. The IST spectra in the insets iii, iv, and v indicate $g=1$ or soliton-type solutions. This shows that TOD has a direct influence on the rogue wave, transforming it towards solitons.

\section{Effect of self-steepening}
\label{L4}

To observe the change arising from the SS term on a rogue wave, we consider the equation:
\begin{eqnarray}
\label{eq6}
i\frac{\partial \psi}{\partial z}-\frac{\beta_2}{2} \frac{\partial^2 \psi}{\partial t^2}+\gamma \,\psi\lvert\psi\rvert^2+is\frac{\partial}{\partial t}(\psi\lvert\psi\rvert^2)=0\textrm{.}
\end{eqnarray}
\begin{figure}[htbp]
\centering
\includegraphics{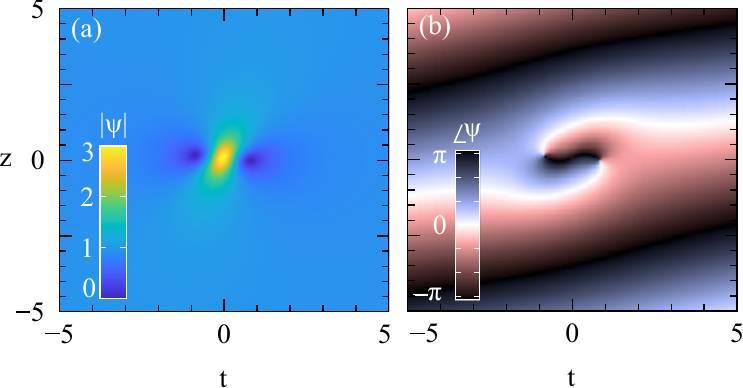} 
\caption{(a) Amplitude and (b) phase profiles of the rogue wave from Eq.~\ref{eq9} with $s=0.2$. It has the maximum amplitude of $3$ with a distorted phase shift of $\pi$ across the peak.}
\label{fig4}
\end{figure}

We note that analytic rogue wave solutions of the derivative nonlinear Schr\"odinger equation, which is similar to Eq.~\ref{eq6}, have been derived and discussed in several previous works \cite{xu2011darboux,liu2019super,chan2014rogue}. These solutions can be converted to the rogue wave solutions of Eq.~\ref{eq6} through a gauge transformation \cite{xu2011darboux}. However, this makes it difficult to observe the solutions in the limit of $s\to0$, as the coefficient $s$ appears in the solution with an inverse relation, making them indeterminate at a small $s$ \cite{xu2011darboux,han2011effect}.

We look for a solution that can be studied without this restriction. We re-formulate the rogue wave solution of the modified NLSE presented in \cite{chen2016chirped,chen2019super} into a simpler form, revealing its clear connection with the NLSE rogue wave solutions. This is given as:
\begin{equation}
\label{eq9}
\begin{aligned}
\psi_{s}(z,t)&= \left(1-\frac{G+ i H z+8 is \tau}{D_s}\right)e^{ i \left[ z \left(1+\frac{1}{2}s ^2\right)-t s+\Phi \right]}\textrm{,}
\end{aligned}
\end{equation}
where $\tau=t-z s$, $\kappa=1+s^2$ and 
\begin{equation*}
\begin{aligned}
 D_s&=D+4 i s  (2 \tau -t)+4 s  \tau  (s  \tau -2 z)\textrm{,}\\
 \Phi &= 2 \tan ^{-1}\left[\frac{4 s  (z s -\tau )}{1+4 \kappa  \left(z^2+\tau ^2\right)}\right]\textrm{.}\\
 \end{aligned}
\end{equation*}
Here, $\beta_2=-1$ and $\gamma=1$, while $s$ can be an arbitrary value. Now, $s$ arises in a way that the solution is not indeterminate as $s\to0$, but rather, it directly reduces to fundamental rogue wave solution in Eq.~\ref{eq3}. Eq.~\ref{eq9} tells exactly how the SS term modifies the fundamental rogue wave solution. The denominator $D_s$ is a complex polynomial which is real for a fundamental rogue wave solution. This solution profile can now be translated to any point on the $z$-$t$ plane following the relations $t=t_1-t_{s}$ and $z= z_1-z_{s}$.

The broken symmetry in the solution is captured in the parameter $s$. The seed modulation acquires a velocity in the transverse direction at the initial stage of the rogue wave development. It gives rise to a new term $\tau=t-zs$ where $s$ introduces the velocity as shown in Fig.~\ref{fig4}. With $s=0$, $\tau=t$, and eliminating the velocity component, Eq.~\ref{eq9} directly reduces to the fundamental rogue wave solution of the NLSE.
 
The effect of $s$ on the rogue wave can be best described by the volume of the rogue wave. The volume is given by \cite{ankiewicz2017multi}:
\begin{equation}
\label{vol}
V =\frac{1}{\sqrt{8\pi}} \int_{-\infty}^{\infty}\int_{-\infty}^{\infty} (I_s)^2 \,dt \,dz\textrm{,}
\end{equation}
where the intensity solution, $I_s=( |\psi_s|^2-1)$, in the integral can be given as:
\begin{equation}
I_s=\frac{8 [1-4 t^2+8 t s  (z+s  \tau )+4 (z-s  \tau ) (z+3 s  \tau )]}{D_g D_g^*}\nonumber \textrm{,}
\end{equation}
where $D_g=1+4 t^2+4 i t s -8 i s  \tau +4 (z-s  \tau )^2$, and $D_g^*$ is its complex conjugate. From this, we obtain the volume $V$ and the rate of volume change with respect to $s$, which are: 
\begin{equation}
V=\frac{1}{1+s^2}, ~\frac{dV}{ds} = -\frac{2 s}{(1+s^2)^2}~~\textrm{.}
\end{equation}
\begin{figure}[htbp]
\begin{center}
\includegraphics{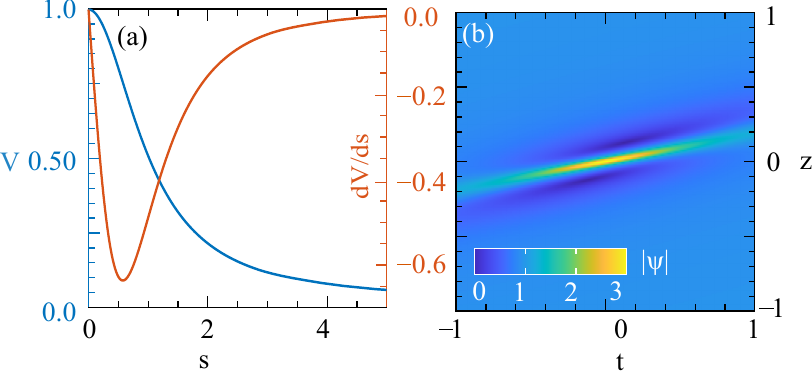}
\caption{(a) Rogue wave volume $V$ (blue) and the rate of volume change $dV/ds$ (red) versus $s$. (b) Onset of a rogue wave departing from its original state to become a soliton when $s=2$.}
\label{volume}
\end{center}
\end{figure}

We subtract the background amplitude $1$ in the integral to avoid having infinite energy from the intensity solution. It is clear that the volume of a rogue wave under the influence of SS effect is varying with the coefficient $s$. A rogue wave is localized both in space ($t$) and time ($z$), and therefore $V$ must be finite.

From Fig.~\ref{volume}(a), we have a finite volume of $1$ when $s=0$, which is the volume of fundamental rogue wave. With an increasing $s$, rogue wave's volume $V$ decreases asymptotically. What this means in the rogue wave behavior is evident in Fig.~\ref{volume}(b). When $s=2$, the rogue wave is stretched in the $t$ dimension creating a highly compressed partly soliton-like profile with $V=0.2$. At a larger value of $s$, this localization is even stronger, and the rogue wave in its most part transforms into a soliton with still a finite $V$ keeping the center-amplitude $3$ unchanged. However, as $s\to\infty$, the rogue wave's volume vanishes and it entirely transform into a soliton. Note that in a real system, $s$ generally remains small, and hence the dynamics remain within a rogue wave partly transformed to a finite volume semi-soliton-type entity instead of a fully transformed soliton. This shows that similar to TOD, the SS effect also has a mechanism to influence the rogue wave to transform it into a soliton.

The rate of change of volume as a function $s$ is also plotted in Fig.~\ref{volume}(a). The fastest change is observed at $s=0.58$ with $dV/ds=-0.65$. Beyond this point, the rate decreases and when $s\gg1$, it approaches $0$. This indicates that at a high value of $s$, the rogue wave is in a soliton-like state.

\subsection{Effects on phase and spectral evolution}

The SS term also induces significant change in the rogue wave's temporal phase and spectral intensity. This term in Eq.~\ref{eq6} is a first-order $t$ derivative of the self-phase modulation implying that instead of a $\pi$ phase shift for NLSE rogue wave, the phase is varying in the transverse $t$ dimension. Due to this, an asymmetric phase term $\Phi$ arises in the NLSE rogue wave solution. From Eq.~\ref{eq9}, this can be given as:
\begin{equation}
\label{eq11}
  e^{i\Phi} =  \frac{D_s^*}{D_s},
\end{equation}
which indicates that the asymmetrical phase profile has its origin in the denominator of the rational solution Eq.~\ref{eq9}. The $t$ varying phase means an instantaneous frequency changes across the envelope. The origin of this change comes from the SS term, which in the Fourier domain becomes $\frac{\partial}{\partial t}(\psi\lvert\psi\rvert^2)=-i\omega (\psi\lvert\psi\rvert^2)$, replacing ${\partial}/{\partial t}=-i\omega$. To deal with this type of asymmetric spectrum, a Heaviside step function is required, which makes the full derivation with arbitrary $s$, $z$, and $t$ rather complicated. We take the Fourier transformation of Eq.~\ref{eq9} at $z=0$:
\begin{eqnarray}
\label{eq15}
\begin{aligned}
F(\omega,s,z=0)&=\frac{1}{\sqrt{2\pi}} \int_{-\infty}^{\infty} \psi_s(z=0,t) e^{i \omega t} \,dt.
\end{aligned}
\end{eqnarray}

For a fundamental rogue wave with $s=0$, the spectrum is:
\begin{eqnarray}
\label{eq16}
\begin{aligned}
F(\omega,s=0,z=0)\,\,&=\sqrt{2 \pi }\left[-e^{-\frac{\left| \omega \right| }{2}} + \delta (\omega )\right] \textrm{,}
\end{aligned}
\end{eqnarray}
where $\delta$ is the Kronecker delta. For $s=0.2$, the expression becomes:
\begin{gather*}
F(\omega,s=0.2,z=0)=5 \sqrt{2 \pi } \delta (1-5 \omega )\\
\,\,\,\,\,\,\,\,\,\,\,\,\,\,-\frac{1}{1521} \Big\{25 e^{-\frac{1}{52} \left(-1+3 \sqrt{3}\right) (-1+5 \omega )} \sqrt{\frac{\pi }{2}} (5+\omega )\\
\,\,\,\,\,\,\,\,\,\,\,\,\,\,\,\,\,\,\,\,\,\,\,\,\times \Big[ a_{1}\theta (0.2-\omega)e^{\frac{3}{26} \sqrt{3} (-1+5 \omega )} +a_{2} \theta (-0.2+\omega )\Big]\Big\}\textrm{,}
\end{gather*}
where $a_{1}=\left(9+14 \sqrt{3}\right)$, $a_{2}=\left(-9+14 \sqrt{3}\right)$, and $\theta$ is a Heaviside step function. Similar spectral expressions can be derived for any values of $s$. For simplicity, we omit the cumbersome mathematical expressions.
\begin{figure}[htbp]
\begin{center}
\includegraphics{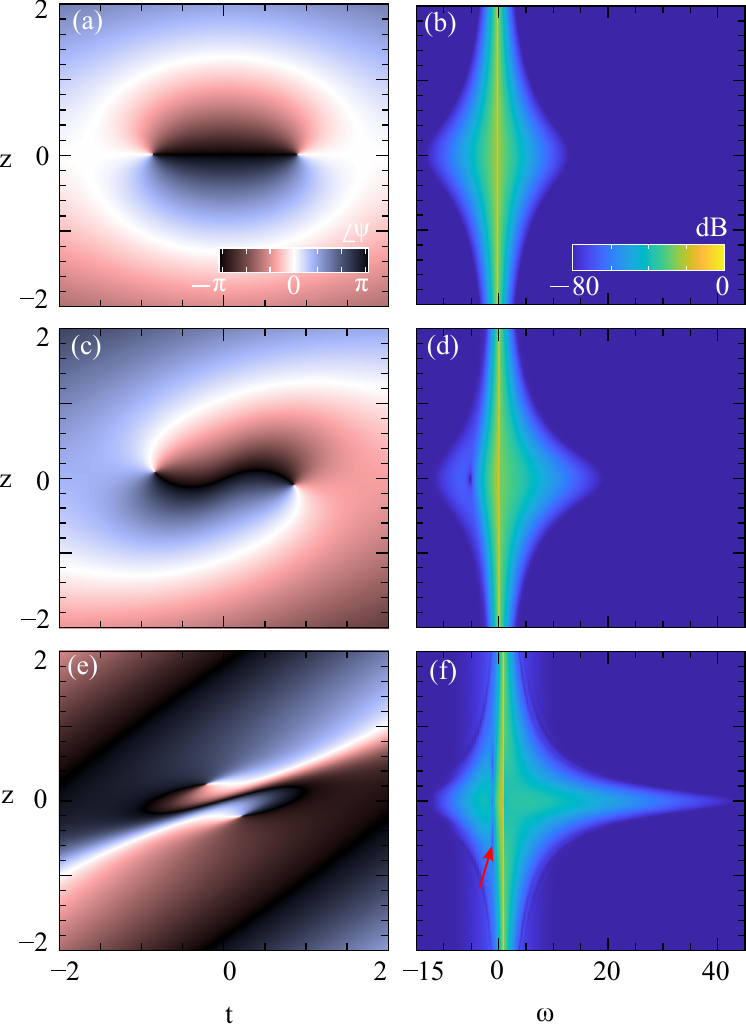} 
\caption{(a) Temporal phase and (b) spectral intensity evolutions of the rogue wave solution, Eq.~\ref{eq9} with $s=0$, $0.2$, and $1$. Note a spectral discontinuity is arise in the spectrum indicated by the red arrow.}
\label{fig8}
\end{center}
\end{figure}

The impact of the SS effect on the phase and spectrum for all $z$ and $t$ is demonstrated in Fig.~\ref{fig8} using the solution presented in Eq.~\ref{eq9}. In Fig.~\ref{fig8}(a), the flat-top $\pi$ phase shift for the fundamental rogue wave with $s=0$ is now distorted in Figs.~\ref{fig8}(c) and \ref{fig8}(e) for $s=0.2$ and $1$, respectively. As a result, the corresponding asymmetrical spectral intensity broadening is observed around $\psi(z=0,t=0)$ which shown in Figs.~\ref{fig8}(d) and \ref{fig8}(f). Note that with $s=0$, the spectrum becomes triangular, which represents a rogue wave, as shown in Fig.~\ref{fig8}(b). With an increasing $s$, the spectral profile broadens asymmetrically towards the blue-side. The maximum phase distortion is seen when $s=1$, which is accompanied by the widest spectral bandwidth.

\subsection{Extended evolution and translation}

\begin{figure}[htbp]
\centering
\includegraphics{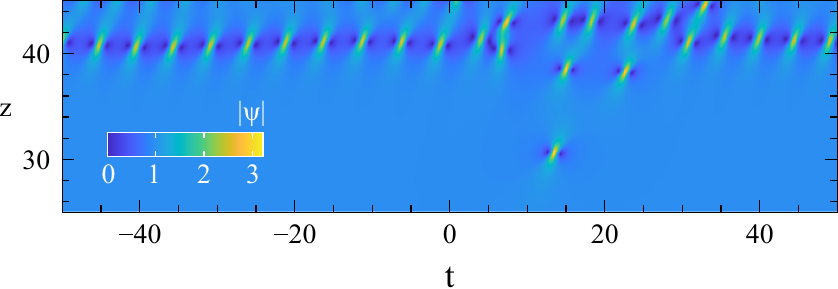} 
\caption{(a) Extended amplitude evolution of a rogue wave with the SS effect when $s=0.2$. The recurring emergence of breathers is achieved also with the same amount of translations as the primary rogue wave. The rogue wave has appeared at a translated distance $t \approx 13$.}
\label{fig9}
\end{figure}

To investigate how the SS effect influences the background after the emergence of the rogue wave, we simulate the fundamental rogue wave for an extended propagation length. The evolution is presented in Fig.~\ref{fig9}. It induces a skewed asymmetry in the recurring breathers after the emergence of the rogue wave. Unlike the effects of TOD, no solitary waves nor dispersive waves are observed in the neighborhood.

Another important aspect of our current observation with the SS effect is, the fundamental rogue wave solution Eq.~\ref{eq3} undergoes a natural translation on the $z$-$t$ plane. We observe that instead of the rogue wave appearing at $t=0$, it emerges translated at $t\approx13$ when $s=0.2$. This indicates that the translational parameter $t_s$ in the analytic solution Eq.~\ref{eq9} is naturally triggered by the finite $s$ in the system. We observe that in the analytic solution, the same amount of translation can be achieve by setting $t_{s}=13$ and $z_s=0$.

\section{Rogue wave self-frequency shift}
\label{L5}

To investigate the RIFS effect on a rogue wave, we employ:
\begin{eqnarray}
\label{eq_raman}
i \frac{\partial \psi}{\partial z}-\frac{\beta_2}{2} \frac{\partial^2 \psi}{\partial t^2}+\gamma \,\psi\lvert\psi\rvert^2-\tau_R \psi \frac{\partial|\psi|^2}{\partial t}=0\textrm{.}
\end{eqnarray}

The Raman term in Eq.~\ref{eq_raman} is a non-Hamiltonian dissipative term \cite{menyuk1993soliton}. Therefore, Eq.~\ref{eq_raman} does not render an analytical solution, and the study is undertaken numerically. Our investigation shows that the RIFS effect does not impact the main rogue wave solution structure in a significant way keeping the amplitude and phase profiles unaltered. However, the impact becomes significant after the emergence of the rogue wave. The RIFS effect mainly induces a steering effect on the rogue wave slowing it down in a transverse direction and gradually transforms it into a soliton.
 
\begin{figure}[htbp]
\centering
\includegraphics{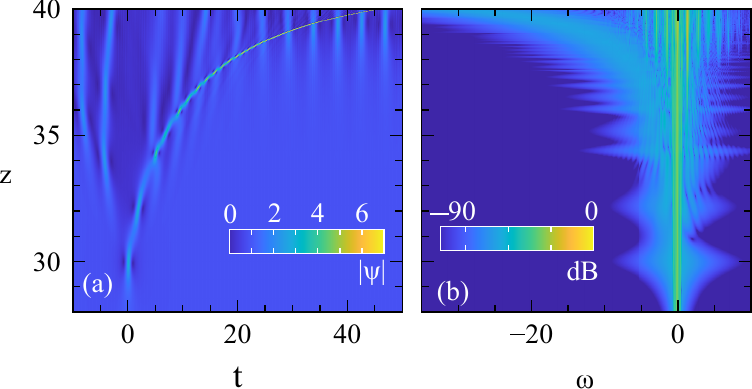} 
\caption{(a) Temporal and (b) spectral intensity evolutions of a fundamental rogue wave under the influence of RIFS effect when $\tau_R=0.012$. The rogue wave appearing at $z=30$ is decelerating along the propagation. A frequency down-shifting is observed in the spectral domain.}
\label{fig12}
\end{figure}

Since the RIFS parameter $\tau_R$ is inversely proportional to the pulse duration, its role becomes greater when the rogue wave reaches the maximum compression point. The dissipative nature of the RIFS effect means that the solution does not preserve the energy. As shown in Fig.~\ref{fig12}, when the rogue wave solution is maximally compressed, its bandwidth is wide enough to amplify the low-frequency components at the expense of the blue-side. At this stage, the rogue wave is no longer robust, but instead it loses energy by generating red spectral components. As the energy dissipation continues, the rogue wave slows down and decelerates. Each time the rogue wave appears with less energy, the compression becomes strong, resulting in a broader spectrum as shown in Fig.~\ref{fig12}(b). Note that, the RIFS-induced radiation in the red-side appears to be small in the first few rogue wave events along $z$. This means that the pulse duration in the first few rogue wave emergences are still not short enough to fully activate the RIFS effect.
\begin{figure}[htbp]
\begin{center}
\includegraphics{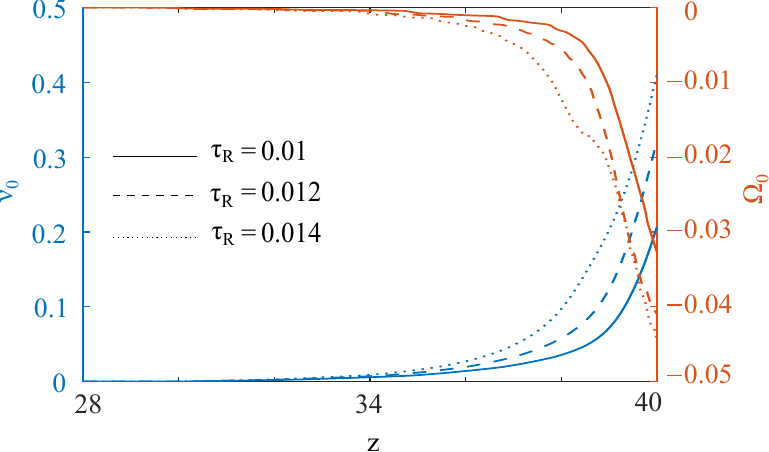} 
\caption{Trajectories of the center of mass of the rogue wave solution in the temporal (blue) and spectral (red) domains under the influence of RIFS effect.}
\label{cm}
\end{center}
\end{figure}

The propagation dynamics can be more effectively described by the progression of the center of mass of rogue wave solution while it is under the influence of RIFS effect. The centers of mass in temporal and spectral amplitudes, $\nu_0$ and $\Omega_0$, are defined as:
\begin{equation}
\nu_0 = \frac{\int_{-\infty}^{\infty}\,t\,|\psi(z,t)|^2 dt}{\int_{-\infty}^{\infty}|\psi(z,t)|^2 dt},\Omega_0 = \frac{\int_{-\infty}^{\infty}\,\omega\,|\psi(z,\omega)|^2 d\omega}{\int_{-\infty}^{\infty}|\psi(z,\omega)|^2 d\omega}
\end{equation}

Figure \ref{cm} shows the trajectories for three different values of the RIFS coefficient, $\tau_R=0.010$, $0.012$, and $0.014$. For high values of $\tau_R$, the magnitude of RIFS also large. With this, the spectrum shifts towards the negative frequency side along the propagation. This results in the rogue wave slowing down in the temporal domain, making an increasingly skewed bow-shaped trail.
\begin{figure}[ht]
\begin{center}
\includegraphics{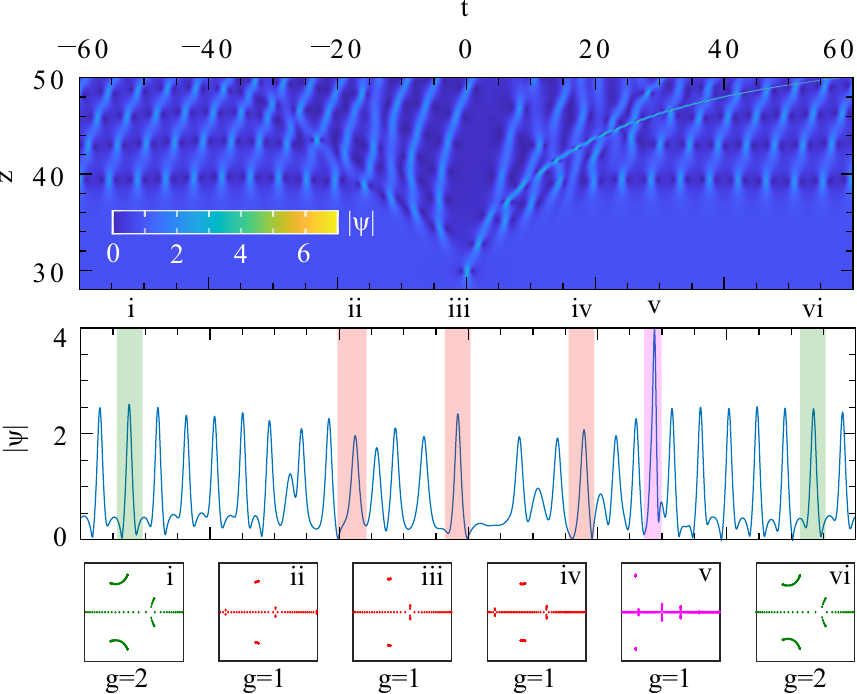}
\caption{(Top panel) Temporal amplitude evolution of a rogue wave solution under the influence of the RIFS effect. A skewed background wave is visible with a various types of wave entities appearing atop. (Mid-panel) an extracted wave envelop at $z=40$ after the emergence of rogue wave. (Bottom panel) IST spectra of various localized wave profiles.}
 \label{fig11}
\end{center}
\end{figure}

We carry out simulation for a longer distance of $z=50$, to investigate the fate of the rogue wave after an extended propagation and its impacts on the neighboring wavefield. This is shown in the top panel of Fig.~\ref{fig11}. We can see that after its appearance at $z=30$, the rogue wave triggers a group of solitons at the central region similar to TOD while its trajectory bends towards the positive $t$ direction. As the propagation progresses, continuous dissipation of energy significantly reduces its transverse velocity. At the same time, the rogue wave gradually becomes a meandering soliton-type entity.

The rogue wave, in its first appearance, extensively changes the nearby background wave. In the top panel of Fig.~\ref{fig11}, the manifestation of a variety of wave groups is clearly visible. To classify types of the wave forms, we conduct an IST analysis. We take a cross-section of the wave envelop at $z=46$ as shown in the mid-panel of Fig.~\ref{fig11}. This is similar to the analysis we carried out earlier in the case of the TOD effect. We select various localized structures from the envelop highlighted in the shaded areas and group them in two separate category according to the genus $g$. The green shaded areas are breather-type, $(g=2)$ localized structures and under the red and magenta-shaded areas are soliton-type where $g=1$. Note that the transformed soliton which is created from the slow moving rogue wave is shaded in magenta.

The localized structures in the (i) and (vi) exhibit three distinct spectral bands, which indicates that they belong to the family of breather-type solutions. The waves in (ii), (iii), (iv), and (v) each have two spectral bands, which corresponds to the soliton-type. Note that in (v), the eigenvalue formation is slightly shifted due to the asymmetry of the profile. This is the soliton shaded under the magenta color that is transformed from the decelerating rogue wave. 

We conclude that the RIFS effect heavily influences the continuous wave background, and triggers a host of other waves such as solitons and breathers. The RIFS effect triggers the formation of a group of low amplitude solitons around the central region where the first rogue wave appeared while the rogue wave itself becomes a relatively high-amplitude slow-moving soliton. The breather-type waves form near the edge of the field away from the center.

\section{Combined effects}
\label{vi}

\begin{figure}[htbp]
\centering
\includegraphics{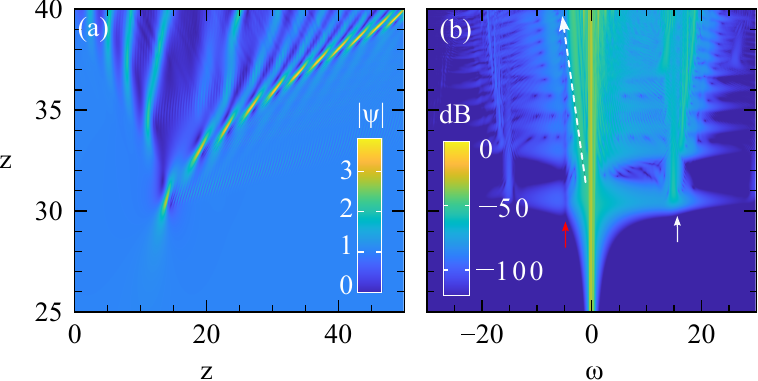} 
\caption{(a) Temporal and (b) spectral intensity evolutions of the fundamental rogue wave when $\epsilon_3=-0.03$, $s=0.2$, and $\tau_R=0.012$. }
\label{fig15}
\end{figure}

Applying all three effects simultaneously leads to Eq.~\ref{eq2}. We solve it numerically to observe their combined influence on the rogue wave solution. The early stage amplitude evolution is shown in Fig.~\ref{fig15} both in temporal and spectral domains, which bears a clear signature of the TOD, SS, and RIFS effects. As shown in Fig.~\ref{fig15}(a), the SS-induced translation shifts the emergence of rogue wave in the positive $t$-direction. Both the TOD and RIFS effect slow down the rogue wave after its appearearance. Their mutual interaction affects the trajectory of rogue wave propagation. This trajectory depends on the magnitude of these two effects.

The TOD onsets the generation of dispersive wave indicated by the white arrow from the first appearance of rogue wave at $z=30$ as shown in Fig.~\ref{fig15}(b). The subsequent emergence of rogue wave in the extended evolution also radiates dispersive waves. The rogue wave compresses to shorter durations and broader spectra further along $z$. The asymmetry in the spectral profile arises due to the SS and RIFS effects. The SS causes the asymmetry towards the blue, while RIFS produces the red-shifted frequency as shown by the dashed long white arrow. The competing effects cause the total spectral profile to be uneven in the transverse directions.

 Note that the spectral discontinuity indicated by the red arrow in the red-side of the spectral profile in Fig.\ref{fig15}(b) is also due to the SS effect. The similar spectral gap can be seen to the left of $\omega=0$ in the analytically obtained spectral profiles in Figs.~\ref{fig8}(d) and \ref{fig8}(f) when $s=0.2$ and $s=1$.

\begin{figure}[htbp]
\centering
\includegraphics{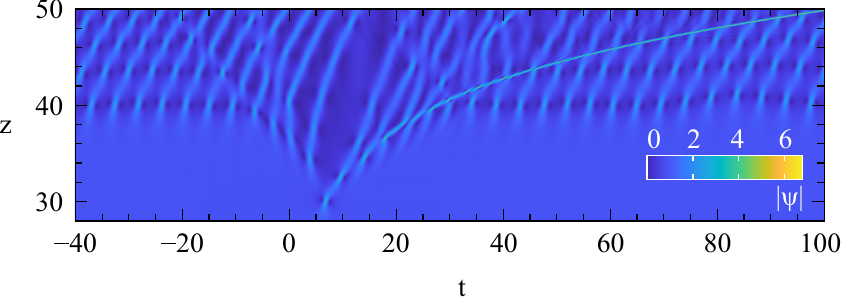} 
\caption{Decelrating rogue wave under the influence of TOD ($\epsilon_3=0.03$), SS ($s=0.1$), and RIFS ($\tau_R=0.008$). It transforms into a soliton.}
\label{fig-soliton}
\end{figure}

Finally, we investigate the long-term evolution of the rogue wave under the combined influence. We observe that at the moment of the rogue wave appearance at $z=30$, it immediately triggers breathers near the edges and a group of low-amplitude solitons around the central part of the propagation field as presented in Fig.~\ref{fig-soliton}. At the same time, due to the RIFS effect, the decelerating rogue wave gradually transforms into a soliton as shown in Fig.~\ref{fig-soliton} with a bent trajectory. Note that as this soliton is propagating on a modulated wave field, the soliton profile is not uniform. It collides with the surrounding breathers-type waves as it advances, and develops ridges on top as seen in Fig.~\ref{fig-soliton}.
\begin{figure}[htbp]
\centering
\includegraphics{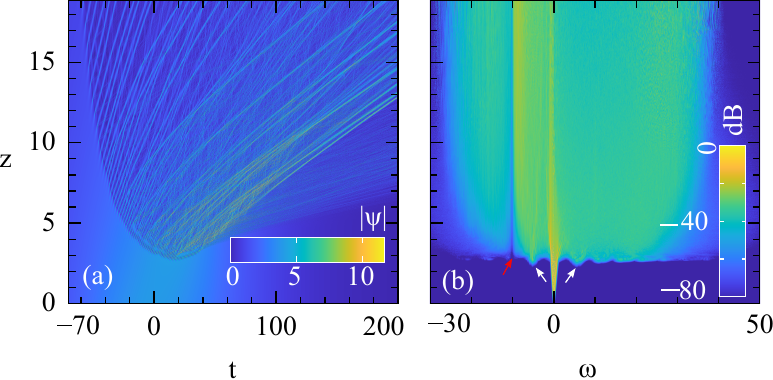}
\caption{(a) Temporal and (b) spectral intensity evolutions showing disintegration of $N=200$ soliton under the influence of TOD ($\epsilon_3=0.03$), SS ($s=0.1$), and RIFS ($\tau_R=0.001$).}
\label{fig-soliton-200}
\end{figure}

To give an example of the implication of the current analysis of how rogue waves can trigger the formation of a large number of solitons, we simulated $N=200$ solitons with TOD ($\epsilon_3=-0.03$), SS ($s=0.1$), and RIFS ($\tau_R=0.001$) using Eq.~\ref{eq2}. The evolution dynamics in the temporal domain can be observed in \ref{fig-soliton-200}(a). Clearly, the conventional perturbation induced soliton-fission involving low number of soliton does not takes place in this case. Instead, the presence of noise among the closely packed high number of solitons triggers the onset of MI which in its initial stage gradually amplifies along the evolution. When the growth of MI is at its peak value, first, it creates a large number of highly compressed rogue wave-type localized structures.

Under the influence of the combined effects, in the final stage, these localized structures delivers a shower of solitons moving towards the positive $t$ direction. These ejected solitons have a close dynamical resemblance to the solitons that we discussed in Fig.~\ref{fig-soliton} under the combined effects of TOD, SS, and RIFS. The evidence of the onset of MI is clearly seen in the spectral profile Fig.~\ref{fig-soliton-200}(b) with the presence of side-lobes indicated by the white arrows. Moreover, a spectral gap is observed next to the left side-lobe that arise from the SS effect shown by the red arrow.

\section{Conclusion}

We studied the dynamical properties of a fundamental rogue wave under the influence of TOD, SS, and RIFS effects. We showed these effects acting on the rogue wave separately as well as simultaneously. We note that all three effects can act individually to generate a bunch of solitons on the background.

After the emergence of rogue wave, TOD can trigger the formation of solitons, breathers, as well as new rogue-wave type formations in its surrounding. Depending on the magnitude of the SS coefficient, a finite volume rogue wave can transform into a soliton. This effect can shift the emerging point of the rogue wave in the transverse dimension. Similarly, under the RIFS effect, a rogue wave generates a group of solitons in the background while the rogue wave itself slows-down via radiating a continuous red shifted frequency and transforms into a soliton. 

When these three effects are applied simultaneously, the rogue wave decelerates and creates an asymmetric bell-shaped spectral profile. However, after an extended simulating propagation, a collection of solitons are created. Simulation of $200$ solitons under the effects of TOD, SS and RIFS, we demonstrated that, indeed the rogue waves that shaped by the noise driven MI finally becomes an ensemble of solitons. In a real system hundreds of rogue wave-type structures are formed due to MI. The purpose of this work is, in the light of the systematic analysis of one of them under the system-perturbation as we carried in this work, explain their collective behavior in a physical system.

We believe these observations present more in-depth understanding of the MI as a nonlinear process where rogue wave-type structures can form. It presents a coherent overview of long-term cw propagation inside optical fiber under the three main higher-order effects. As a representative case, the observation revealed how MI leads to the formation of rogue waves that finally transforms into a large collection of solitons. Also, a comprehensive understanding of the first-order rogue wave evolution in the presence of the higher-order effects may shed light on propagation dynamics of higher-order rogue wave solutions. 

\section*{Acknowledgements}
This work is supported by Ministry of Education, Singapore (2019-T2-2-026). A.C. and M.B. acknowledge financial support from the Nanyang Technological University, NAP-SUG.

\bibliographystyle{unsrt}

\end{document}